# Analysis Traceability and Provenance for HEP

**Jetendr Shamdasani, Richard McClatchey, Andrew Branson and Zsolt Kovács**

University of the West of England, Coldharbour Lane, Bristol, BS16 1QY

jet@cern.ch

**Abstract**. This paper presents the use of the CRISTAL software in the N4U project. CRISTAL was used to create a set of provenance aware analysis tools for the Neuroscience domain. This paper advocates that the approach taken in N4U to build the analysis suite is sufficiently generic to be able to be applied to the HEP domain. A mapping to the PROV model for provenance interoperability is also presented and how this can be applied to the HEP domain for the interoperability of HEP analyses.

## 1. Introduction

In complex data analyses it is increasingly important to capture information about the usage of data sets in addition to their preservation over time in order to ensure reproducibility of results, to verify the work of others and to ensure appropriate conditions data have been used for specific analyses. This so-called provenance data in the computer science world is defined as the history or derivation of a data (or process) artifact [1]. Many scientific workflow based studies are beginning to realize the benefit of capturing the provenance of their data and the activities used to process, transform and carry out studies on that data. This is especially true in scientific disciplines where the collection of data through experiment is costly and/or difficult to reproduce and where that data needs to be preserved over time. With the increase in importance of provenance data many different models have emerged to capture provenance such as the PROV [2] or the OPM [3] models. However, these are more for interoperability of provenance information and do not focus on the capture of provenance related data. There is a clear and emerging requirement for systems to handle the provenance of data over extended timescales with an emphasis on preserving the analysis procedures themselves and the environment in which the analyses were conducted alongside the processed data sets.

A provenance data system that has been built in house at CERN since early 2000 is called CRISTAL [4]. CRISTAL was used to capture the provenance resulting from the design and construction of the CMS ECAL detector over the period 2002-2010. The CRISTAL Kernel (V3.0) has now been launched as open source under the LGPL (V3) licensing scheme and it is available for use by the wider communities including teams involved in the offline analysis of physics data, whether at CMS or other experiments. In addition, in the EC funded neuGRID [5] and N4U [6] projects the original developers have been using CRISTAL to capture the provenance of analyses for neuroscientists running complex pipelines of algorithms in the study of biomarkers for the onset of Alzheimer's disease. In this paper this application is presented with a focus on how its approach can be customized for use in the high energy physics data analysis community at large. The main focus of this is a set of analysis tools (persistency, browsing/querying, visualizing and analysis tracking services) which together with a generic analysis model backend can be used to capture the information required to support complex analyses.

CRISTAL is based on what is known as a "Description Driven" approach where objects (or Items in CRISTAL terminology) are described instead of programmed at runtime. It is a flexible system which has been used outside of research in industry [7]. It has a strong pedigree and can be potentially useful for the tracking of HEP analyses. It is provenance enabled by design which means that Items created in CRISTAL are tracked, their history and the changes recorded to these items and are tracked and stored as extra metadata about the object. The object or Item itself is not thrown away and is stored in addition; therefore different versions can be loaded up at run-time if required. CRISTAL, as mentioned previously has been the "core" of the N4U analysis suite.

The purpose of this paper is to present the so called "Analysis Suite" from the N4U project and how it could be adapted to the provenance enabled analysis capture for physics analyses. The next section presents the internal CRISTAL model, section 3 presents the Analysis Tools developed for N4U, section 4 shows how provenance was applied in N4U, section 5 shows the top level PROV export model for analyses, section 6 shows how the tools and methods developed for N4U could be used for the HEP domain and finally section 7 concludes the paper with directions for future work.

## 2. Experimental Analysis in N4U

This work has been conducted in the context of the EC funded N4U (neuGRID for You) project which is a follow on from the original neuGRID project [5, 6]. The goal of the N4U project has been to create a set of tools to make the running of neuroscience experiments on the neuGRID infrastructure more user friendly. In the case of N4U these experiments are image analysis jobs that run in a distributed manner on the GRID. One main difference from the neuGRID project and N4U is the idea of *provenance*. In the case of the N4U project it was chosen to build its analysis software on top of the CRISTAL system. CRISTAL (now known as CRISTAL-iSE [8]) is a fully provenance enabled system in which all data is tracked and nothing is thrown away. In the case of the N4U project it was chosen to initially map the internal CRISTAL provenance model onto the so-called W7 model [9]. In the case of the N4U project the analysis tools were able to capture the following information:

- **who** ran an analysis, this is a user name,
- for **what** purpose, what their analysis is supposed to achieve,
- **when** they ran it this is a timestamp which denotes when it started and when it finished,
- **where** it was run this is GRID and Cloud related information,
- **which** datasets and algorithms were used to create and run their analyses,
- **how** it was executed, this more detailed infrastructure information
- and lastly **why** the analysis was run, this is a justification from the user.

Within CRISTAL provenance is event based, this means that once an activity completes (in a workflow for example) it generates an event. During this event generation, provenance information is captured. Figure 1 below shows the CRISTAL model that was developed in the N4U project. In figure 1 the blue elements are CRISTAL Items (akin to an object in object oriented design) and the pink elements are known as Item data, which are inputs or outputs from Items. This is the internal model used to capture the Analysis information from N4U. The key thing to note here is that the model is very abstract. This was done intentionally so that it may be extended and reused for other domains that require analyses to be performed. The Analysis Item in this model is able to handle any set of input Parameters which come from Pipelines (synonymous with Workflows). There is a one-to-one mapping between a Pipeline and a Data Element (which is a single file/image from a dataset); this information is captured in the Analysis Element Item which is later sent to the Grid for processing, which generates a set of Results for the end user. For further description of this model and its usage in the project please see [10].

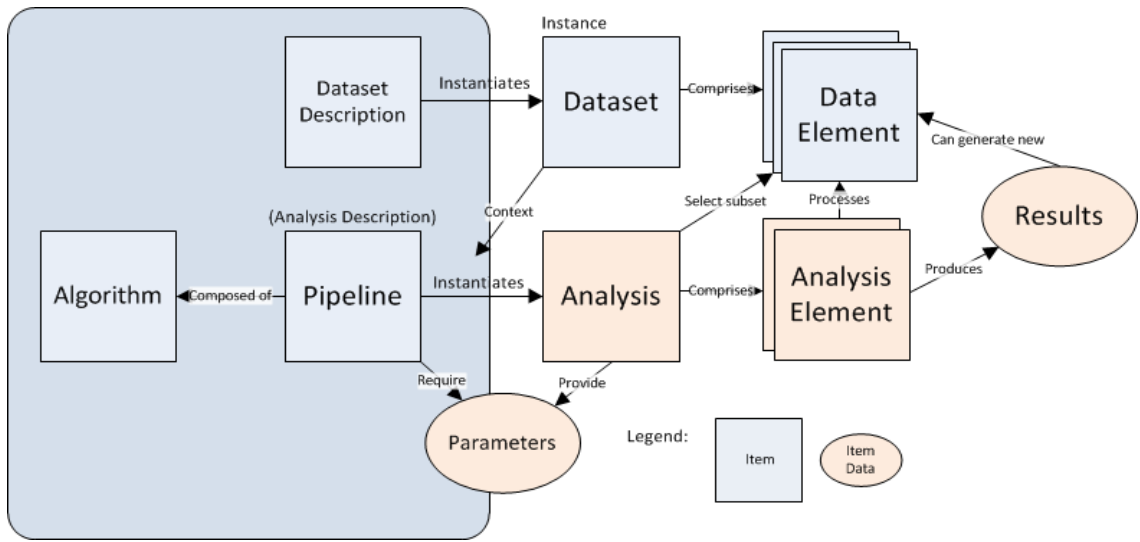

Figure 1 : CRISTAL N4U Model

## 3. Analysis Tools in N4U

Within the project, besides implementing a model for the capture of analyses, there were a set of tools developed for users to interact with the CRISTAL based analysis backend. Figure 2 shows these tools and how they interact with each other.

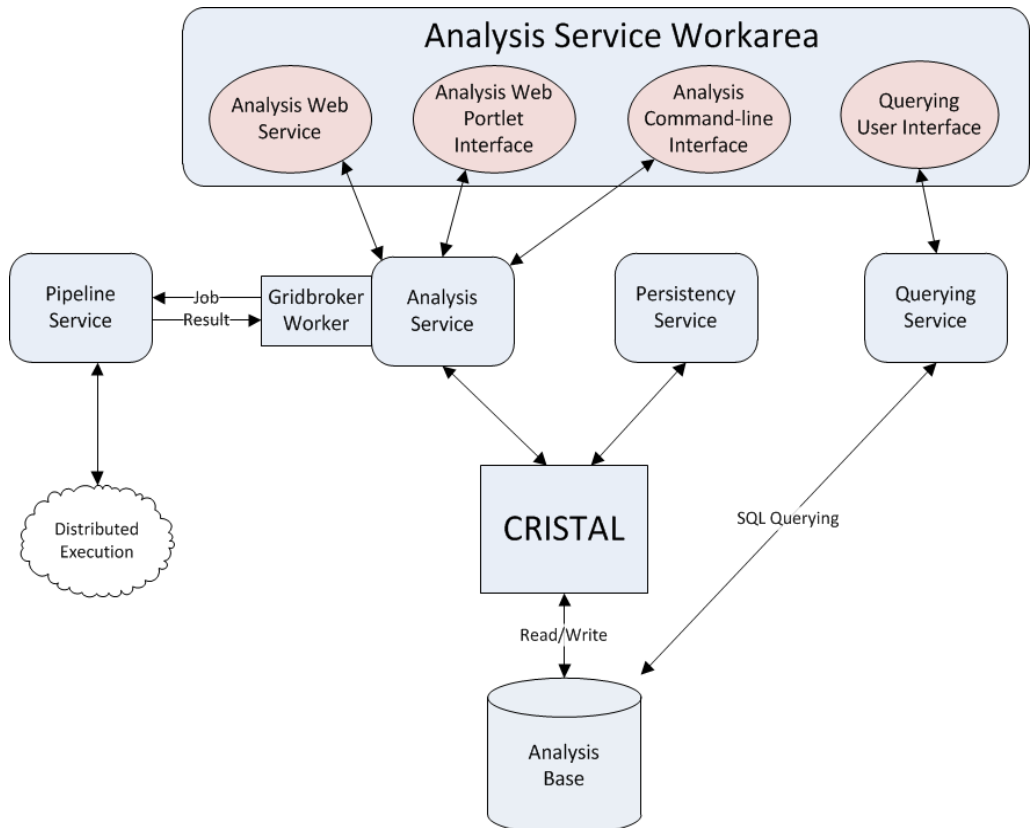

Figure 2 : N4U Analysis Architecture

In Figure 2 there is a large Analysis Service Workarea. This workarea consists of four interfaces that allow a user to interact with the various backend services. These are the following:

- **Analysis Web Service** – This is a SOAP [11] based web service that allows users in N4U to programmatically access the analysis suite of tools. This web service has the ability to submit analyses to the N4U infrastructure and to query past analyses.
- **Analysis Web Portlet Interface** – This is a portlet based UI component which allows users to see the status of their jobs and/or to submit analyses. This interface is simple and is meant for novice users to be able to act with the backend analysis services. The portlet interface is based on Liferay [12] and Vaadin [13].
- **Analysis Command Line Interface** – This is akin to a bash prompt, this is for users to be able to access and create scripts to create their own analyses.
- **Querying User Interface** – This is used to view and query data available in the Analysis Backend.

As well as the frontend interfaces present in the Analysis Service Workarea, there are several backend services which can be used alone (with no interface, programmatical or otherwise). These are the Analysis Service, which is the main "core" of the system and exposes the analysis objects from the CRISTAL server, the Persistency Service which is used to import and export data from the CRISTAL backend in a raw XML format and finally the Querying service which is mainly used to transform queries from the querying UI into SQL queries to be able to access the information stored in the Analysis Base.

It should be noted that there is a Grid Broker element present; this is used by the Analysis Service to transform users' analyses into a language that is understandable by the N4U Pipeline Service which is used to submit jobs to the N4U GRID infrastructure. The Grid Broker is also capable of unmarshalling the results returned by the Pipeline Service into the common analysis CRISTAL Item based language that is used by the Analysis tools. The provenance is collected along the way in a seamless manner which is completely transparent to the user. The next section discusses this approach.

**4. Provenance in N4U**

The key factor in N4U is that it is using the CRISTAL system which, as outlined above, is provenance enabled. This means that everything is tracked and nothing is discarded. In figure 3 the basic idea is presented of how the provenance approach was implemented in N4U. The key notion in the figure is the concept of "Provenance Enabled Objects" in what is known as the Analysis Base, shown at the foot of the diagram [14].

These "Provenance Enabled Objects" are CRISTAL Items that have provenance in N4U. These include Subject Metadata, Data Structure Definitions, Images (and their metadata), Pipelines (and their metadata), Queries and finally Analyses. By these objects containing provenance, it is meant that all changes to these objects are tracked over time and (re-)use therefore they contain an audit trail or history of what has happened to them to make them materialize in their current state. For example, if one is to consider a pipeline, these can come in different versions with different configuration parameters. When a pipeline is changed or updated and re-uploaded to the N4U Analysis Workarea the changes to it are tracked, along with who did the changes and why they did it. The same is true for the other provenance enabled objects. Later versions of these objects can be loaded at any time (if required) by a user. Users can also see the provenance of other Analyses performed in the past (such as who executed an analysis or the result of the analysis) and they can also try and recreate the analysis with similar parameters.

The detailed operation of the Analysis Service is best understood with a practical example. Consider the case where a clinician wishes to conduct a new analysis. Her first step would be to compile a selection of data from the datasets which are available to her. To do this she would log into the Analysis Service Workarea (top of figure 3) and interact with the Querying Service through its user interface to find data that possesses the particular properties she is looking for.

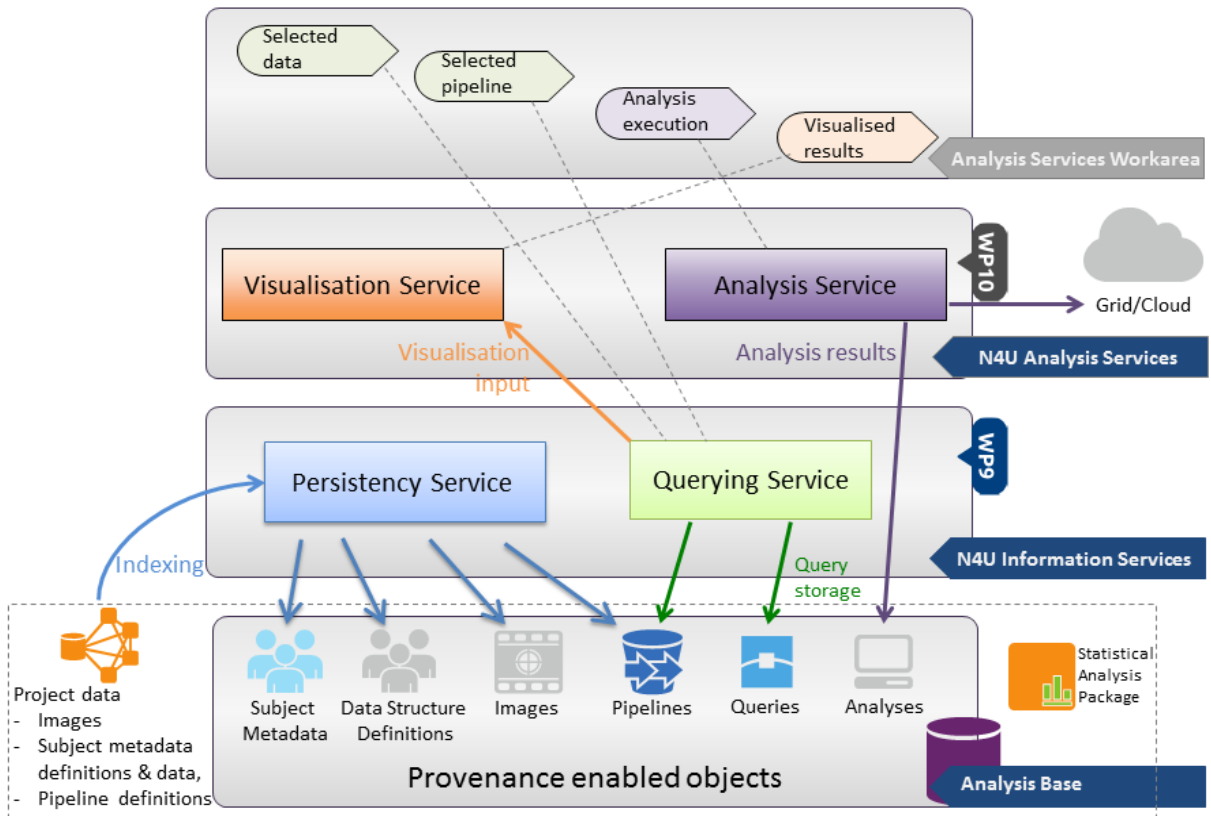

Figure 3 : The N4U Virtual Laboratory

She submits her constraints, which are passed as a query to the Querying Service. The Querying Service then queries the Analysis Base which would return a list of dataset properties and locations which meet her constraints. The Querying Service interface would then display this list to the clinician to approve. Once the user is satisfied with her dataset selection she combines it with a pipeline specification to create her Analysis. To do this she would need to use the Analysis Service (a frontend) to search CRISTAL for existing algorithms that she can use to create a new pipeline or select a pre-defined pipeline. Command line utilities are provided to aid in the creation of a pipeline by connecting different algorithms together as steps. The completed pipeline will have a dataset associated with it and once this pipeline is ready it will be run on each element of the dataset by CRISTAL.

The pipeline will be sent to CRISTAL which will orchestrate the input pipeline (see figure 3) using a Job Broker and the N4U Pipeline Service. Single activities from the input workflow will be sent to the Pipeline Service as a single job using the pipeline API. Once the job has completed, the result will be returned to CRISTAL. Here CRISTAL will extract and store provenance information for this job. This information will contain factors such as the time taken for execution, and whether the job completed successfully. It will store this information internally in its own data model. It will also post this information to the Analysis Base so that this crucial provenance information is accessible by the Querying Service. This loop of sending jobs and receiving the result will continue until the workflow is complete. Once this workflow has completed CRISTAL will once more generate provenance information and store this provenance for the entire workflow in its own internal data store and the Analysis Base. The final result of the completed workflow/pipeline will be presented to the user for evaluation. A link to the completed result in the form of a LFN (a GRID location) will be stored in the Analysis Base. The clinician now has a permanently logged record (provenance data) of her analysis including the datasets and (versions of) algorithms she has invoked, the data captured during the execution of her analysis and the final outcome and data returned by her analyses. These provenance elements may also have associated annotation that she has added to provide further knowledge of her analysis that she or others could consult at a later time to re-run, refine or verify her analysis. It is clear such functionality is generally applicable across many scientific disciplines, including HEP.

**5. Mapping to PROV**

As well as tracking the provenance in real-time of the analyses performed by neuroscientists, the system is able to perform an export to a "standard" provenance format, in this case PROV. PROV was chosen instead of the OPM since it is now a W3C standard for Provenance and is quickly emerging to be the choice for *provenance interoperability*. It consists of three main super classes (Entities, Activities and Agents) and seven main relationships in between these classes and recursive relationships between the classes themselves. For a full description and tutorial on PROV please see [15]. Figure 4 shows the top level PROV model used in N4U.

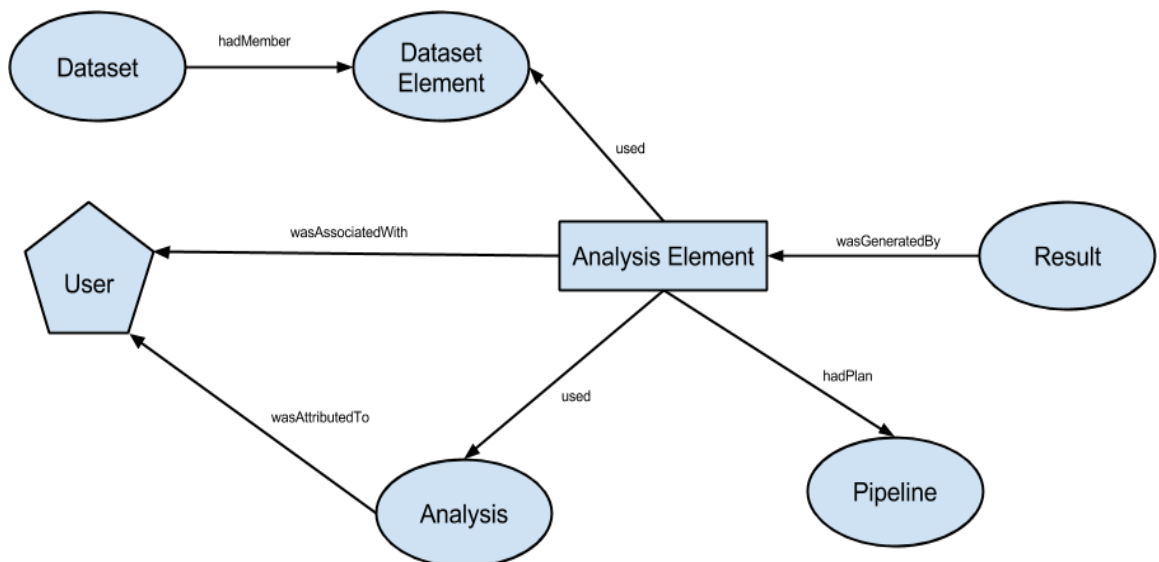

Figure 4 : Top level PROV analysis model

The ovals represent *Entities*, the hexagons are *Agents* and the rectangles are *Activities*. This is using the PROV graphical notation. The main concepts have been kept the same as the CRISTAL internal data model for N4U except that now the Analysis Element (instead of being in Item) is an *Activity* in PROV. This decision was made due to the fact that the Analysis Element itself is the main execution object and does not contain provenance, but instead is a combination of a dataset and a pipeline. The main top level relationships that are used are *used, wasAssociatedWith*, *wasAttributedTo* and *wasGeneratedBy*. In this version of the model, the cyclical relationships are implied (e.g. an Entity *wasDerivedFrom* another Entity). Besides the core top level PROV relationships there are sub relationships used such as *hadPlan* (to show that a Pipeline is a plan of execution) and *hadMember* to demonstrate that a DatasetElement is a member of the collection Dataset.

This export to PROV was a key part of the project, since it demonstrates interoperability with standards. Theoretically a user can now export their analysis to PROV and use a standard PROV based analysis tool such as [16] to visualize the provenance information. Besides the export being able to be used for the interoperability, it allows Neuroscientists to see how other neuroscientists in other experiments have used their own analysis tools. For example of a scientist has used Taverna [17] which also supports a PROV export, a comparison can be done from the export from the N4U analysis suite to see how the analyses compare e.g. to look at the path taken to achieve a similar result. This provenance trace not only shows the trace of execution, but the decisions taken by different scientists to achieve the same goals. Therefore, this information is useful from a scientific stand point since users can compare not only the final result, but the method used to achieve these results. The PROV model presented here is a high level model which is able to capture information from *any* analysis therefore, it is implementation independent. This provenance is not just N4U specific but it is *Analysis* specific. It is felt that this high level description of analysis provenance can be reused in any domain (along with the CRISTAL model). The next section discusses this approach.

## 6. Mapping to HEP

Figure 1 shows the internal CRISTAL model and how it captures Analysis information. CRISTAL as stated previously was originally developed at CERN and used to track the construction of the CMS ECAL detector. Many of the concepts from the original CMS application have been applied to the world of Neuroscience analysis. The major concept that was taken forward was the *traceability* aspect. This traceability was extended during the course of neuGRID and N4U to track the analyses performed by neuroscientists. One of the expansions that were performed was the GridBroker concept in figure 2. This broker was essential since it made CRISTAL compatible with the Grid and therefore the distributed execution of algorithms. This has allowed CRISTAL to become a fully functional execution environment for the development (with the aid of the Analysis tools), tracking and execution of Analyses. The key thing to note here is that this "broker" component can be replaced to work with other infrastructures such as the Cloud. The tools, as mentioned previously have been written in a generic manner. This means that they can be reused for the domain of HEP.

In section 4 there were a set of "Provenance Enabled Objects" presented. This means that all these objects or *Items* in CRISTAL are all provenance aware. Therefore, they have full traceability; every change that happens to these objects is recorded since they first come into being. A user can look at these and see how they came to be in their current state. Which is useful, not just for the domain of Neuroscience but it can be applied to the domain of HEP as well. For example if a HEP scientist wants to see what datasets (and the version there of) was used with a certain algorithm they can perform a simple query within CRISTAL allowing a user to extract this information instantly.

The PROV model in section 5 is generic; this is a top-level capture of the domain of Analyses. This is very generic and the export can be performed from inside of the Analysis suite. This model is meant to be extended can be adapted to any domain that uses Analyses (e.g. HEP). Exporting an analysis to PROV has the main benefit that it is a W3C standard. Therefore, other systems are capable of performing a PROV export as well. This means that using a PROV visualization tool such as [16] a user can load two separate analyses into it and see the provenance from these analyses. This is very relevant for the HEP domain since different experiments use different analysis tools. Therefore, exporting the provenance to a common format allows users to see what experiments were performed in the past and to be able to compare visually what the difference were to achieve a (similar) result. This is also a first step towards the reproducibility of experiments since information such as the workflows and datasets are captured as metadata. Therefore an analysis could be recreated at a later date in a different system. This allows for a harmonization of the mapping and transformation of analysis information. This is a much more realistic and simpler approach than making all the experiments use the same analysis software which may not fit all their requirements.

Another advantage of the PROV export is that it is stored as text (RDF and XML); it can be preserved for use at a later date and can be viewed by any text editor if required. This allows for the preservation of this information since text formats are standard now in the computing domain. This means that the preservability of the information is high and the storage footprint is low. Therefore, it is easily accessible at a later date and can be understood by reading the PROV graph (in any notation that it is stored in). Since this information is structured (as a graph) it is more readable than a log file, which can be in varying formats. As mentioned previously, PROV is now a W3C standard, therefore, this allows increased openness. Which is important for physicists since openness of the results and the processes used to achieve these results are crucial to aspects like education and reproducibility.

## 7. Conclusion and Future Work

This paper has presented some experience of using the CRISTAL provenance management tool to create analyses in the neuroscience domain. The functionalities of the tools in the N4U analysis suite have been presented and discussed through examples of how scientists carry out their many and varied analyses. The core of this paper was to outline how the N4U Analysis Tools (which are provenance aware) could be used in the HEP domain to support physicists' complex data analyses. Our next steps will be to gather real analysis data from one of the HEP experiments at CERN and to use our N4U analysis approach to be able to capture the provenance of HEP analyses and thus support ongoing HEP analyses. This work will enable users in the HEP community to be able to see how other user's analyses may have performed, to reproduce and verify the results of those analyses and to be able to capture the provenance of these analyses and preserve this over time. For a more detailed discussion of the analysis tools the interested reader is referred to [6] and to [4] for a more detailed discussion of CRISTAL and its provenance approach. CRISTAL is now open source is available for general use by the public as a multi-purpose provenance data base (http://cristal-ise.github.io).

**Acknowledgements**

The authors wish to highlight the support of their home institutes and acknowledge funding from the European Union Seventh Framework Programme (FP7/2007-2013) under grant agreement n. 211714 ("neuGRID") and n. 283562 ("neuGRID for users") with special thanks to the N4U consortium.

**References**

[1] Simmhan, Yogesh L, et al. "A survey of data provenance in e-science." *ACM Sigmod Record* 34.3 (2005): 31-36.

[2] PROV - http://www.w3.org/TR/prov-overview/

[3] Moreau, Luc, et al. "The open provenance model core specification (v1. 1)."*Future Generation Computer Systems* 27.6 (2011): 743-756.

[4] Branson, Andrew, et al. "CRISTAL: A practical study in designing systems to cope with change." *Information Systems* 42 (2014): 139-152.

[5] Redolfi, Alberto, et al. "Grid infrastructures for computational neuroscience: the neuGRID example." *Future Neurology* 4.6 (2009): 703-722.

[6] McClatchey, Richard, et al. "Traceability and Provenance in Big Data Medical Systems." *CBMS (2015).*

[7] Shamdasani, Jetendr, et al. "CRISTAL-ISE: provenance applied in industry." *ICEIS* (2014).

[8] McClatchey, Richard et al. "Designing Traceability into Big Data Systems." *ICT: BDCS* (2015).

[9] Ram, Sudha, and Jun Liu. "A New Perspective on Semantics of Data Provenance." *SWPM* 526 (2009).

[10] McClatchey, Richard, et al. "Providing traceability for neuroimaging analyses." *International journal of medical informatics* 82.9 (2013): 882-894.

[11] SOAP - http://www.w3.org/TR/soap

[12] Liferay - http://www.liferay.com/

[13] Vaadin - https://vaadin.com/home

[14] Munir, Kamran et al. " Development of a Large-Scale Neuro-Imaging and Clinical Variables Data Atlas in the neuGRID4You (N4U) Project." *Journal of Biomedical Informatics*. July 2015. DOI:10.1016/j.jbi.2015.08.004.

[15] Moreau, Luc, and Groth, Paul. "Provenance: An introduction to prov." *Synthesis Lectures on the Semantic Web: Theory and Technology* 3.4 (2013): 1-129.

[16] Hoekstra, Rinke, and Groth, Paul. "PROV-O-Viz-Understanding the Role of Activities in Provenance." *IPAW* (2014).

[17] Wolstencroft, Katherine, et al. "The Taverna workflow suite: designing and executing workflows of Web Services on the desktop, web or in the cloud." *Nucleic acids research* (2013).